\documentclass[%
 reprint,
 amsmath,amssymb,
 aps,
 pra,
]{revtex4-2}

\usepackage{graphicx}
\usepackage{dcolumn}
\usepackage{bm}
\usepackage{hyperref}

\usepackage[ruled,linesnumbered,noline]{algorithm2e}
\usepackage[braket,qm]{qcircuit}
\usepackage[english]{babel}

\newcommand{\dirac}[1] { | #1 \rangle }

\newtheorem{definition}{Definition}
\newtheorem{theorem}{Theorem}
\newtheorem{proposition}{Proposition}
\newtheorem{lemma}{Lemma}
\newtheorem{corollary}{Corollary}
\newtheorem{example}{Example}
\newtheorem{conjecture}{Conjecture}

\newcommand{\bdfn}{\begin{definition} \begin{rm}}
\newcommand{\edfn}{\hfill $\Box$ \end{rm} \end{definition}}
\newcommand{\bthm}{\begin{theorem} \begin{rm}}
\newcommand{\ethm}{\hfill $\Box$ \end{rm} \end{theorem}}
\newcommand{\bprop}{\begin{proposition} \begin{rm}}
\newcommand{\eprop}{\hfill $\Box$ \end{rm} \end{proposition}}
\newcommand{\bcor}{\begin{corollary}\begin{rm}}
\newcommand{\ecor}{\hfill $\Box$ \end{rm} \end{corollary}}
\newcommand{\balg}{\begin{algorithm} \begin{rm}}
\newcommand{\efact}{\hfill $\Box$ \end{rm} \end{nota}}
\newcommand{\bex}{\begin{example} \begin{rm}}
\newcommand{\eex}{\hfill $\Box$ \end{rm} \end{example}}
\newcommand{\bconj}{\begin{conjecture} \begin{rm}}
\newcommand{\econj}{\hfill $\Box$ \end{rm} \end{conjecture}}
\newenvironment{proof}%
  {\medskip \noindent {\bf Proof:}\hspace{1ex}}%
  {\hfill $\Box$}

\begin{document}

\preprint{APS/123-QED}

\title{Quantum invariants for the graph isomorphism problem}

\author{Hern\'an I. de la Cruz}
 \email{HernanIndibil.Cruz@uclm.es}
\author{Fernando L. Pelayo}
 \email{FernandoL.Pelayo@uclm.es}
\author{Vicente Pascual}
 \email{vpfuniversity@gmail.com}
\author{Jose J. Paulet}
 \email{jpaulet@ucm.es}
\author{Fernando Cuartero}
 \email{Fernando.Cuartero@uclm.es}
\affiliation{%
 Universidad de Castilla - La Mancha, Spain
}%

\author{Luis Llana}
 \email{llana@sip.ucm.es}
\affiliation{%
 Universidad Complutense de Madrid, Spain
}%

\author{Mauro Mezzini}
 \email{mauro.mezzini@uniroma3.it}
\affiliation{%
 Roma Tre University, Italy
}%

\date{\today}

\begin{abstract}
Graph Isomorphism is such an important problem in computer science, that it has been widely studied over the last decades. It is well known that it belongs to NP class, but is not NP-complete. It is thought to be of comparable difficulty to integer factorisation. The best known proved algorithm to solve this problem in general, was proposed by L\'aszl\'o Babai and Eugene Luks in 1983. 

Recently, there has been some research in the topic by using quantum computing, that also leads the present piece of research. In fact, we present a quantum computing algorithm that defines an invariant over Graph Isomorphism characterisation. This quantum algorithm is able to distinguish more non-isomorphic graphs than most of the known invariants so far. The proof of correctness and some hints illustrating the extent and reason of the improvement are also included in this paper.
\end{abstract}

\maketitle


\section{Introduction}

The problem of whether two graphs are isomorphic, that is, determining the existence of any bijection between the vertices set of both graphs that preserves the graph structure, has been a great challenge for the scientific community for a long time. Both theoretical and computational aspects related to this problem have been extensively discussed in the \hyphenation{scien-ti-fic} literature in the last decades. The problem of graph isomorphism is not only interesting from a theoretical point of view but also from the point of view of the computational complexity since the class of complexity to which this problem belongs remains uncertain. It is not yet determined whether the graph isomorphism problem can be considered as either a $P$-problem or a $NP$-complete problem or it belongs to a different category. It was even described as a disease in 1977~\cite{ReCo77}, due to the fact that every one who has studied it felts into a obsession about it.

As a consequence, many authors have addressed this problem so proposing different approaches to solve it but, so far, an efficient algorithm to conclude in any possible case, whether or not two given graphs are isomorphic is not known. Until now, the best known algorithm to generally solve the isomorphism problem was proposed by  L\'aszl\'o Babai and Eugene Luks in~\cite{BL83}. It holds a computational complexity of $O(2^{\sqrt{n \log n}})$ where $n$ is the number of vertices in the graph.

It is well known the fact that two graphs having an identical characteristic polynomial is an invariant, i.e., it is a necessary (but no sufficient) condition for them to be isomorphic. Two graphs sharing the same characteristic polynomials are called {\it isospectral} graphs and they have been extensively studied in the literature. There is also an algorithm~\cite{BGY84} that relies on this fact, for the case of eigenvalues holding some restrictions.

Although, as mentioned before, an efficient algorithm for the general case has not been found yet, there are some algorithms that can efficiently solve the problem in specific cases, like Nauty~\cite{McKa81}, the first efficient method that was able to deal with large graphs using branching techniques. Later, Saucy~\cite{Saucy04} proposed the use of disperse data structures. Other improvements were added by Bliss~\cite{Bliss07} and Conauto~\cite{Conauto09} introducing the use of a novel branching technique.

In 2015, L\'aszl\'o Babai reported the discovery of a quasi-polynomial time algorithm based on the one hand on a divide and conquer scheme and on the other hand on a smart use of both group theory and combinatorics, so decomposing the problem in the search of patterns based on {\it Johnson Graphs}, leading to an algorithm that solves the problem in time $\displaystyle O(2^{\log(n)^{c}})$ for some constant $c$. This result, published in \cite{Babai16} has not been confirmed yet.

Currently the lack of general efficient methods is not a hard problem since we can even say that the graph isomorphism problem has been solved from a practical point of view. There are two main reasons supporting this claim. First, there is a significant amount of graphs for which there is an efficient procedure and, second, there are some heuristic algorithms that provide suitable solutions.

Most of these efficient techniques, including Babai approach, are based on combinatorics, group theory and branching techniques. A good classification can be found in \cite{Conte04} where the authors rely on some other papers. 
In \cite{Hallgren2003916} the authors provide the first general understanding of the Hidden Subgroup Problem, HSP over non Abelian groups. As a positive case  when $H$ is an arbitrary subgroup of $G$ and $H^G$ is the largest subgroup of $H$ that is normal in G, it is true with high probability that $H^G$ is uniquely determined in an efficient way, that is, within polynomial time. However, this generalisation to non Abelian case does not solve efficiently the graph isomorphism problem.
In the paper \cite{PhysRevA.89.022342} the authors move the GI problem into a combinatorial optimization problem to be solved by a Quantum Approximate Optimization Algorithm, QAOA. Nevertheless, there is still no full consensus over the complexity of a general quantum adiabatic algorithm on the related literature.

This paper is structured as follows: Section 2 presents the common pattern for codifying graphs. Section 3 presents one main contribution of this paper as a variation scheme for codifying graphs. Section 4 is devoted to present on detail the algorithm for characterizing graphs which represents the core of the paper. Conclusion and future work section finishes the paper.

\section{Graphs codification}

A graph is a collection of nodes, some pairs of them connected by an edge. Formally, an undirected/non directed graph is a pair $G=(V,E)$, where $V=\{1, 2, \ldots, n\}$ $( |V| = n )$ is the finite set of vertices/nodes, and $E$ is a set of \emph{edges} where each edge is a set of exactly two vertices belonging from $V$. $E\subseteq V \times V$ is the set of edges where loops are not included, i.e., without edges in the form $(x,x)$ for any node $x$.

Given two graphs $G_1 = (V_1,E_1)$ and $G_2 = (V_2,E_2)$, we say that they are isomorphic if there
exists a bijection $b : V_1 \to V_2$ such that $(x,y) \in E_1 \iff (b(x),b(y)) \in E_2$.

Quantum states entanglement technique to model a non-directed graph was first conceived by Hein et al in \cite{Hein06, Hein04}, where the entanglement states in an $n$-particles system are supporting a mathematical description of a graph where each vertex corresponds to a qubit and edges represent interactions between them. 

In order to do this, 
the former system of $n$ qubits must be in a superposition of base states $\dirac{0}$ and $\dirac{1}$, i.e., in $\dirac{+}$ state. Afterwards, each edge $u =(a,b) \in E$ is encoded by a controlled $Z$ quantum gate over the qubits $a$ and $b$ no matter which acts as control and which is controlled.

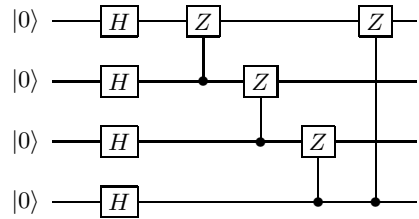
\begin{figure}[htbp]
\begin{displaymath}
    \Qcircuit @C=1em @R=1.2em {
    \lstick{\ket{0}} & \qw & \gate{H} &\qw &\gate{Z}  &\qw       &\qw       &\gate{Z} &\qw\\
    \lstick{\ket{0}} & \qw & \gate{H} &\qw &\ctrl{-1} &\gate{Z}  &\qw       &\qw      &\qw\\
    \lstick{\ket{0}} & \qw & \gate{H} &\qw &\qw       &\ctrl{-1} &\gate{Z}  &\qw      &\qw\\
    \lstick{\ket{0}} & \qw & \gate{H} &\qw &\qw       &\qw       &\ctrl{-1} &\ctrl{-3}&\qw  \\    
  }
\end{displaymath}
  \caption{Example. Graph circuit $C_4$}
  \label{F1}
\end{figure}

Figure \ref{F1} shows the circuit modelling the $C_4$ graph, i.e. graph $G=(V, E)$, where $V = \{1, 2, 3, 4\}$ and the set of edges is $E=\{(1,2), (2,3), (3,4), (1,4)\}$. This corresponds to a circuit graph of 4 nodes (i.e. $C_4$).

\begin{figure*}[htbp]
    \centering
    \includegraphics[width=15cm]{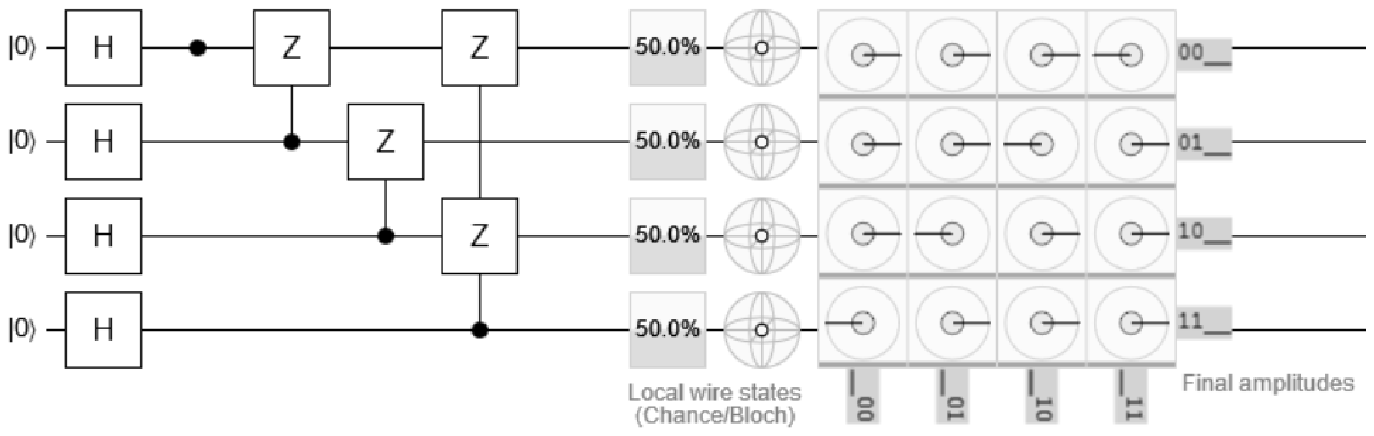}
  \caption{Example 2. Circuit for $\ket{M_1}$}
  \label{F2}
\end{figure*}

Right hand side of figure \ref{F2} shows graphically the amplitudes of the components of the Hilbert space basis on Quirk-QS (Quantum Simulator), which can also be explicitly represented by

\begin{align*}
  \dirac{C_1} = \frac{1}{4} (\dirac{0000} + \dirac{0001} + \dirac{0010} - \dirac{0011} \\
  + \dirac{0100} + \dirac{0101} - \dirac{0110} + \dirac{0111} \\
  + \dirac{1000} - \dirac{1001} + \dirac{1010} + \dirac{1011} \\
  - \dirac{1100} + \dirac{1101} + \dirac{1110} + \dirac{1111})
\end{align*}

According to this codification $C_4$ graph is represented by the amplitudes of the entangled states, in fact the absolute value of their modules meet in all of them, where for the components corresponding to the 4 edges the sign is negative whereas for the remainder 12 the sign is positive.

This encoding scheme provides us with quite interesting features from which we are trying to get the most. In \cite{Zhao16} the authors use this fact to study properties of the graphs leveraging on the performance increasing associated with quantum computing. Furthermore, this quantum encoding allows an efficient handling so as to overcome classical algorithms for common tasks in this field of research. In particular, first section of the given reference is focused on how it could be used for dealing with graph isomorphism problem, nevertheless they finally stated that this idea is not enough to properly characterize graph isomorphism, sic: {\it $\ldots$ it is tempting to believe that different graphs necessarily lead to states that are entangled in different ways. This is not the case, however, as graph states generated from different graphs may be equivalent up to local operations.}

In a quite similar way the problem is faced in \cite{Mills19} where this encoding scheme is taken to provide invariants for the graph isomorphism problem. Two main modifications over the proposal by Zhao el al. are now made. The first is just an aesthetic issue in the sense that it uses the adjacency matrix which under the restrictions of non-directed graph without any self loop, is equivalent to just one of the triangles out of the main diagonal (i.e. either row $>$ column, or, row $<$ column). Each column in the triangle is encoded in a column of quantum gates, the 0 of the diagonal of the adjacency matrix is encoded by a control. These quantum gates are $Z$ when there is an edge and $I$dentity when there is no edge.

For the previous example of $C_4$ graph, its adjacency matrix can be represented by three ways where being one of them $M_1$ is the following

\[M_1 \ = \ \begin{pmatrix}0 & 1 & 0 & 1\\
1 & 0 & 1 & 0\\
0 & 1 & 0 & 1\\
1 & 0 & 1 & 0\end{pmatrix}\]

Figure \ref{F2} shows the corresponding codification of $\ket{M_1}$ on Quirk-QS. 

A noticeable contribution is made by Mills et al. in \cite{Mills19} where weighted graphs can be represented by means of the use of controlled {\it some root of $Z$} gates, instead of the former Pauli $Z$ gates which mean a rotation of $\pi$ radians on Z axe.

Making use of these adds-on that paper is focused (one more paper) on searching for invariants for graph isomorphism characterisation and some other related problems. There are many classical invariants for this purpose as the characteristic polynomial associated to the adjacency matrix for which two isomorphic graphs share the same invariant but the contrary is not true. Invariants searching is an interesting motto to finally find one which works both sides, i.e. a bijection between the quotient set of graphs isomorphisms-modulo and, the set of values for that invariant provided that, of course,  this would be possible.

In the cited paper \cite{Mills19} they make use of Wigner's functions in this searching for invariants task. These functions allow characterising the set of reached states (see \cite{Mills17}). Experiments for this task has been carried out over the IBM Quantum Experience computers so showing that they are able to discriminate a number of invariants for graph isomorphism problem.

\section{Motivation and invariants}

As previously stated, Mills et al. in \cite{Mills19} proposed using controlled Z quantum gates to model a graph. The Z gate performs a rotation of $pi$ radians on Z axe. But our proposal is to use gates with a smaller angle for rotations, as in that paper do when dealing with weighted graphs.

In the following example, we use the T gate, which corresponds to a rotation of $\pi/4$ radians. In general, we will use the general parameterised $R_z (\varphi)$ gate, when $\varphi$ is the angle of rotation around the Z axis, with $T=R_z(\frac{\pi}{4})$. We make use of this gate in order to show the codification of the $C_4$ graph, the circuit of 4 nodes.

\begin{figure*}[htbp]
    \centering
    \includegraphics[width=15cm]{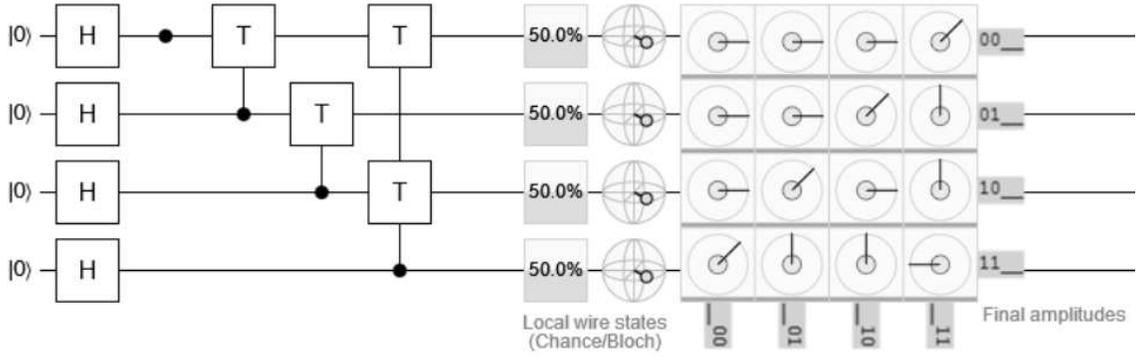}
  \caption{Example 2. Circuit for $\ket{M_1}$ with $T$ gates}
  \label{F3}
\end{figure*}

As previously indicated, we believe that these gates can be used beyond the proposal of Mills et al. for weighted graphs. And first, we can see how using $T$ gates instead of $Z$ ones when encoding $C_4$ graph, the new quantum circuit on Quirk-QS becomes as depicted in figure \ref{F3}.

The detailed description of the corresponding quantum state is

\begin{align*}
  \dirac{M_1} = \frac{1}{4} (\dirac{0000} + \dirac{0001} + \dirac{0010} + e^{i \ \pi/4} \dirac{0011} \\
  + \dirac{0100} + \dirac{0101} + e^{i \ \pi/4} \dirac{0110} + e^{i \ \pi/2} \dirac{0111} \\
  + \dirac{1000} + e^{i \ \pi/4} \dirac{1001} + \dirac{1010} + e^{i \ \pi/2} \dirac{1011} \\
  + e^{i \ \pi/4} \dirac{1100} + e^{i \ \pi/2}\dirac{1101} + e^{i \ \pi/2}\dirac{1110} + e^{i \ \pi}\dirac{1111})
\end{align*}

Going deeper on the amplitudes graphically depicted in the referred Quirk-QS snapshot, it is easy being aware that variations between positive and negative values that appeared when using just $Z$ gates become now into rotations on the phase component of each of the amplitudes. Therefore, 7 out of the 16 components of the whole quantum state keep a 0 radians phase component; 4 out of 16 hold a $\pi/4$ radians phase component (the rotation provided by once operating $T$ quantum gate); another 4 (out of 16) appear as rotated $\pi/2$ radians (twice $T$ applying) and the last of the 16 has a component of $\pi$ radians which is equivalent to 4 times $T$ operations. 

This distribution of amplitudes corresponding to the 16 elements of the base in the Hilbert Space for this state can be observed
in detail in the table \ref{tab:m1-amplitudes}.

\begin{table}[htbp]
\begin{tabular}{ccc}
\hline \hline
Basis vector   & Phase   & N. of $\pi/4$ rotations \\ \hline
$\dirac{0000}$ & 0       & 0                       \\
$\dirac{0001}$ & 0       & 0                       \\
$\dirac{0010}$ & 0       & 0                       \\
$\dirac{0011}$ & $\pi/4$ & 1                       \\
$\dirac{0100}$ & 0       & 0                       \\
$\dirac{0101}$ & 0       & 0                       \\
$\dirac{0110}$ & $\pi/4$ & 1                       \\
$\dirac{0111}$ & $\pi/2$ & 2                       \\
$\dirac{1000}$ & 0       & 0                       \\
$\dirac{1001}$ & $\pi/4$ & 1                       \\
$\dirac{1010}$ & 0       & 0                       \\
$\dirac{1011}$ & $\pi/2$ & 2                       \\
$\dirac{1100}$ & $\pi/4$ & 1                       \\
$\dirac{1101}$ & $\pi/2$ & 2                       \\
$\dirac{1110}$ & $\pi/2$ & 2                       \\
$\dirac{1111}$ & $\pi$   & 4                       \\
\hline \hline
\end{tabular}
\caption{M1 state amplitude analysis}
\label{tab:m1-amplitudes}
\end{table}

This way of representing graphs can be straightforward generalized to any graph such way that once the quantum gate has been chosen, the amplitude associated to each element of the base corresponds to a whole {\it multiple} of the rotation provided by the referred quantum gate.

As we have chosen $T$ quantum gates $T = Z^{1/4}$, 8 different tentative rotations can appear at most (more than just 2 as provided by $Z$ q-gate). We believe that having overcome this restriction release us from suffering the problem stated by Zhao et al., i.e. lacking from the discrimination power required for some pairs of non-isomorphic graphs that just looking at their entanglement properties seemed wrongly being so. We have taken the route of increasing the accuracy of the quantum gates used for this sake.

Elaborating this requires knowing how the rotations on the amplitude phase are made, which comes from the fact that each element of the Hilbert's space base corresponds to a subgraph from the one under consideration. This subgraph is formed after taking the nodes having a $1$ within the sequence of bits that defines it on the basis vector. On the example of $C_4$ graph we are using, the element of the base $\dirac{1011}$ represents the subgraph including nodes $\{1,3,4\}$, which contains two edges ($(1,4)$ and $(3,4)$). Finally each controlled $T$ quantum gate acts over the two qubits related with the edge so generating an extra rotation of $\pi/4$ radians on every amplitude in which both qubits meet on value $1$. This way makes the amplitudes adding up all the rotations corresponding to the edges included in the given subgraph.

It is immediate to discover that the phase angle of the amplitude for the base vector is   $\pi/2$ coming from twice rotating $\pi/4$. This fact also shows that the phase angle of the amplitude can also provide the number of edges corresponding to the associated subgraph.

The above consideration is always true as coming straightforward from the effect of the controlled phase displacement quantum gate. This gate performs such displacement over all the amplitudes in which both qubits meet on $1$, so adding up all these amplitudes therefore also computing the number of edges.

We are now on condition of stating, as conjecture, a tentative invariant for the sake of addressing the graph isomorphism problem. We start with the following

\medskip

\bprop \label{Prop1} Given two graphs $G=(V,E)$ and $G'=(V',E')$ with $n$ nodes each one, they are isomorphic if and only if there exists a bijection $f$

\[ f : \{1 \ldots n\} \longrightarrow \{1 \ldots n \}\]

holding that for every nodes subset of $V$, $S= \{x_1, x_2, \ldots x_j \} \subseteq V$ there is a corresponding subset of $V'$, $S' = \{x_{f(1)}, x_{f(2)}, \ldots x_{f(j)} \} \subseteq V'$ such that the corresponding subgraphs over those subsets have the same number of edges.

\proof

If there exists an isomorphism between $G$ and $G'$ then there is a permutation on nodes in $V$ such that edges on $E$ correspond to edges on $E'$. In fact, this permutation defines $f$ bijection required to conform $S'$ properly as it covers all the subsets of nodes. Therefore it is an invariant.

The other way round is also true, for this it is enough considering $f$ bijection acting over sets $S$ and $S'$ with two nodes each, this defines the permutation characterizing the isomorphism since all the possible edges are defined on both $S$ and $S'$. It is immediate that edges exist together with the correspondant on the other set, i.e. isomorphism's definition is held.
\eprop

Proposition \ref{Prop1} states that from the phase amplitudes of our codification all the possible subgraphs are reachable, so this, the previously referred cite of Zhao et al. \cite{Zhao16} is refuted, in fact assuming the use of quantum gates corresponding to small enough rotations on $Z$ axis (with an appropriate value for the $k$ parameter referred at the beginning of this section), the underlying quantum modeling of graphs that relies on the entanglement on the qubits phases is enough to characterize different graphs. 

Nevertheless, the number of subgraphs for a graph of $n$ nodes is $2^n$ what could seem that we move from a given problem to another one with higher complexity than the former's one. Fortunately this is not this way in the end, since leveraging on the quantum algorithm allows us to classify all the $2^n$ subgraphs by their number of edges, which also will conform a new invariant for our problem.

\section{New invariants}

The presented codification provides a full characterization of different graphs by means of their amplitudes. Nodes permutation can order those amplitudes even when all of them share the same module because an order on their phases can be established. This order ranges from $0$ to the total number of edges of the graph.

There are some drawbacks arising when the accuracy required for the quantum gates at modelling edges must guarantee that the last/biggest amplitude (corresponding to the whole graph, so including the maximum number of edges) is required being smaller than $2\pi$, i.e. for our example of $C_4$ graph a quantum $T$ gate is enough as $T$ allows up to 7 rotations ($\pi/4$ radians) before reaching $2\pi$ radians. It is immediate that this 4-edges graph can easily be modelled. In fact, all the graphs with 4 or less nodes can be modelled. Notice that a 1000 nodes graph could have up to $1000 \cdot 999 /2$ edges for the complete graph, in which case a quantum gate like $Z^{1/(1024 \cdot 512)}$ (for accuracy reason) is required for amplitudes not to reach $2\pi$ in the phase component.

On practice, this imposes hard conditions for designing such very accurate quantum gates. Nevertheless, this is not new since e.g. Shor's algorithm also requires this sort of accuracy demands even when all the amplitudes being smaller than $2\pi$ radians on the phase angle is not required (because modular arithmetic could be used).


Another drawback is the impossibility to direct access the amplitudes of the quantum state. Instead, measuring the system (making the quantum state to collapse) is needed. Quirk shows in many examples that the probability of measuring either $\dirac{0}$ or $\dirac{1}$ is 50\% in all the qubits of the system which, of course, lacks of any practical information.

\begin{figure}[htbp]
  \begin{displaymath}
    \resizebox{.9\hsize}{!}{\Qcircuit @C=1em @R=1.2em {
      \lstick{\ket{\psi}} & \qw {/}  & \qw &\gate{U^{2^0}} &\gate{U^{2^1}}  &  \qw & \cdots &  & \qw  & \gate{U^{2^{n-1}}}     & \qw &\qw &\qw & \qw & \qw\\
      \lstick{\ket{0}}    & \qw      & \gate{H}     &\ctrl{-1} & \qw &\qw  &   \cdots & &\qw   &\qw   &\qw & \multigate{3}{QFT^\dagger}  &\qw & \meter & \qw\\
      \lstick{\ket{0}} & \qw & \gate{H} &\qw &\ctrl{-2}       &\qw &\cdots & &\qw    &\qw      &\qw & \ghost{QFT^\dagger} & \qw & \meter & \qw\\
      \lstick{\vdots}  & &&&&&& \lstick{\vdots} &&&&&& \lstick{\vdots} \\
      \lstick{\ket{0}} & \qw & \gate{H} &\qw &\qw       &\qw       & \cdots & &\qw  &\ctrl{-4} &\qw &\ghost{QFT^\dagger} & \qw & \meter & \qw  \\    
  }}
  \end{displaymath}
  \caption{Quantum phase estimation circuit}
  \label{QPEinShor}
\end{figure}
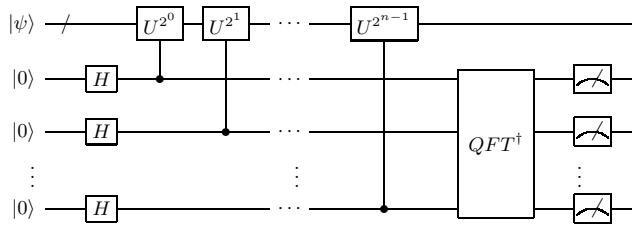

Therefore, a system transformation to make phase angles becoming amplitude modules in order to be aware of these differences is required. The quantum phase estimation, QPE for short, algorithm \cite{QPE_95} has been chosen for this purpose. This algorithm estimates the phase of the eigen-vector associated to a given quantum gate $U$. It is also present on the quantum computing part of Shor's algorithm \cite{Shor99}. This quantum algorithm moves the differences on the phase in the graph codification proposed in our model to a difference on the amplitudes that could be measured on practice. An sketched image of QPE into Shor algorithm can be seen in figure \ref{QPEinShor}.

\subsection{Quantum Phase Estimation Algorithm applying}

\begin{figure*}[htbp]
    \centering
    \includegraphics[width=15cm]{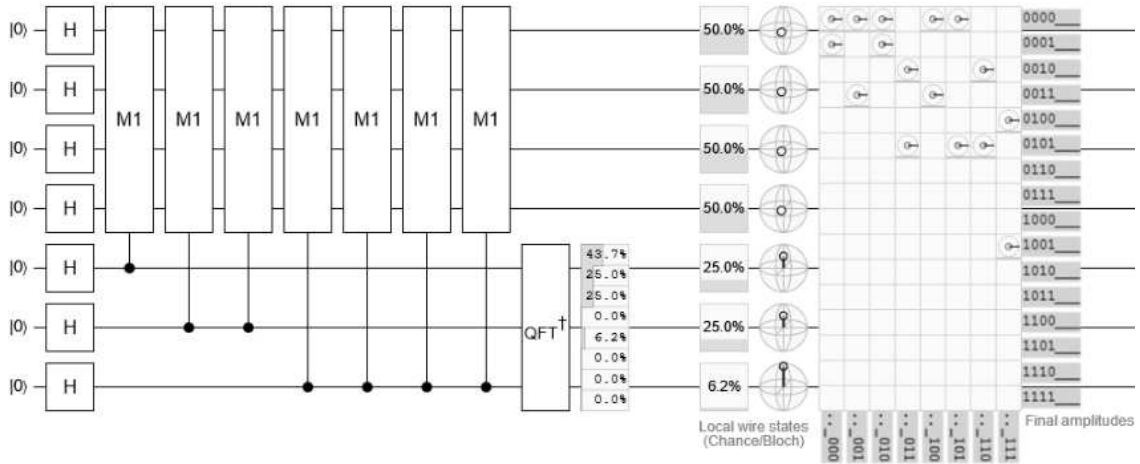}
  \caption{Evaluation of graph $M_1$}
  \label{EF1}
\end{figure*}

The following example is used in order to graphically show how does QPE Algorithm work. Let $M_1$ be the piece of quantum algorithm corresponding to the $C_4$ graph encoded on Figure \ref{F3} that is taken as oracle. The quantum phase algorithm over $M_1$ circuit is shown in Figure \ref{EF1}.

Applying the quantum phase algorithm over $M_1$ offers a set made of the 128 amplitudes of the Hilbert space. Only 16 among them are non-zero so corresponding to the 16 subgraphs. The non-zero components of the new quantum state are:

\begin{align*}
    \dirac{M_1} = \frac{1}{4} (\dirac{0000000} + \dirac{0000001} +  \dirac{0000010} + \dirac{0000100} \\
    + \dirac{000101} + \dirac{0001000} +  \dirac{0001010} + \dirac{0010011} \\
    + \dirac{0010110} + \dirac{0011001} + \dirac{0011100} + \dirac{0100111} \\
    + \dirac{0101011} + \dirac{0101101} + \dirac{0101110} + \dirac{1001111})
\end{align*}

These non-zero amplitudes associated to base vectors should be understood as follows, qubits 1:3 binary encode the number of edges of the subgraph, besides remainder qubits 4:7 describe the presence/absence (1/0) of nodes within the corresponding subgraph. Next 3 paragraphs elaborate on detail, by means of our case study of $C_4$, how we are numbering qubits as well as how we encode and decode each digit of the {\it ket} in the Hilbert base.

Right hand side of figure \ref{EF1} shows all these amplitudes where all of them but 16 lack of any phase ($0$ radians) whereas the rest of them  (positive amplitudes) are those in which we are interested.

Let us notice that first 3 qubits corresponds with control qubits (downside of Quirk circuit) which are computing the number of edges in the subgraphs. Top line in Quirk has 5 positive amplitudes out of 8 which corresponds with the base elements $\dirac{0000000}$, $\dirac{0000001}$, $\dirac{0000010}$,  $\dirac{0000100}$ and $\dirac{0000101}$, where first three bits (1:3) $000$ mean $0$ edges, and, the last four bits (4:7) those in charge of identifying the nodes involved, i.e. $\emptyset$, $\{1\}$, $\{2\}$, $\{3\}$, and $\{1,3\}$ respectively. The line behind the previous one stands for the remainder subgraphs without any edge but with nodes $\{4\}$ and $\{2,4\}$.

For the rest it is easy to see that base element $\dirac{0010011}$ represents the single-edge subgraph $\{1,2\}$, the following single edge subgraphs are $\{2,3\}$, $\{1,4\}$ and $\{3,4\}$. The remainder non-zero amplitudes corresponds with 4 two edges subgraphs with three nodes each and the last refers the complete four-edges with four nodes subgraph $C_4$.

Going deeper on this, we state that just a single measurement on the control qubits provides us with so much information in the following way. One by one throws probabilities $25\%$, $25\%$ and $6.25\%$ of being $\dirac{1}$. If we perform three measures at once we get the probability of reaching each of the values of the   measurement (binary encoded) multiplied by the total number of subgraphs (16 in this case) representing the amount of subgraphs with such number of edges. In the example of $C_4$ graph it includes 7 subgraphs with no edges, 4 subgraphs with one edge, 4 subgraphs with two edges and the only four edges subgraph, i.e., itself.

\[\begin{array}{ccc}
   \#(edges)  & \% Probability & \#(subgraphs)  \\
    0 & 43.75 & 7 \\
    1 & 25.0 & 4 \\
    2 & 25.0 & 4 \\
    3 & 0.0 & 0 \\
    4 & 6.25 & 1 \\
\end{array}\]

\begin{figure*}[htbp]
    \centering
    \includegraphics[width=15cm]{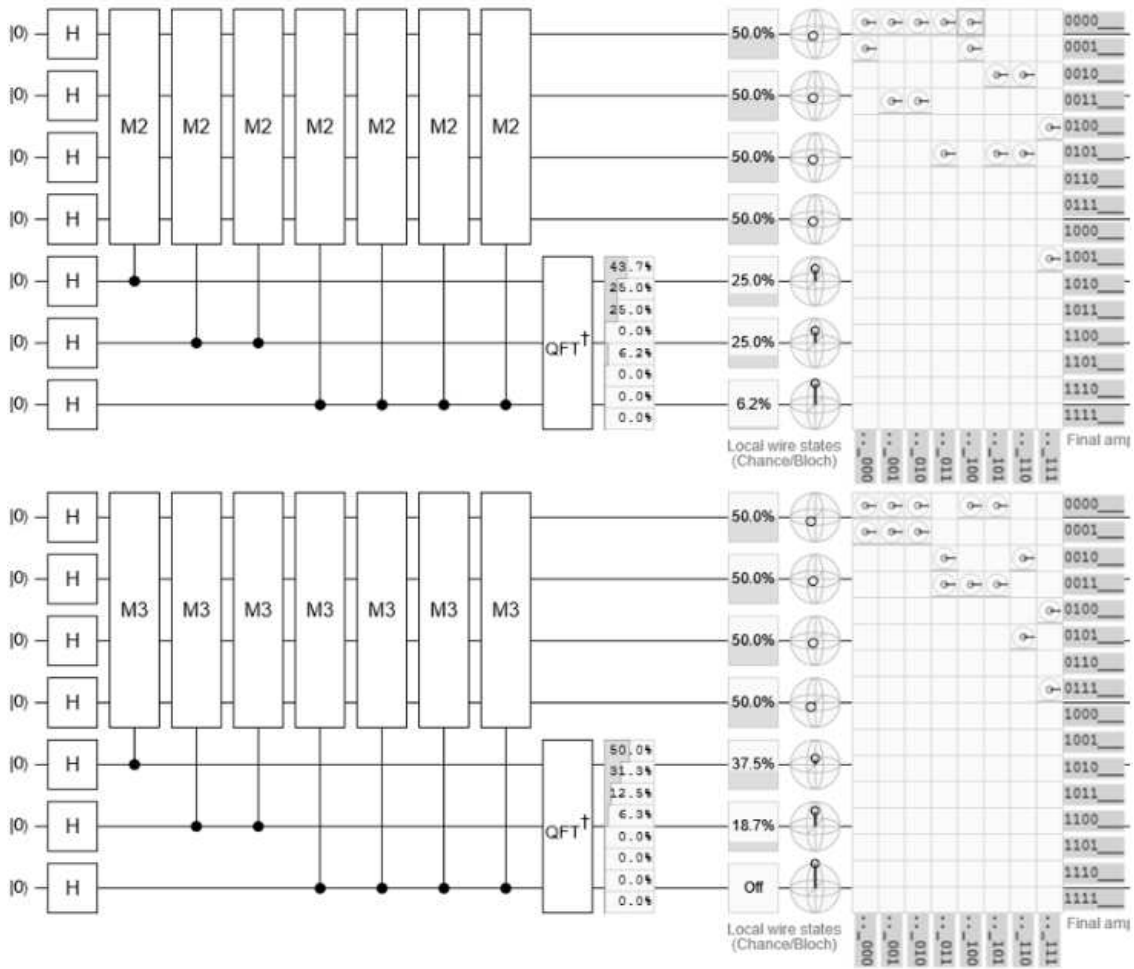}
  \caption{Evaluation of graphs $M_2$ and $M_3$}
  \label{EF2}
\end{figure*}

For the sake of better illustrating the way it works, let us consider the following adjacency matrices of graphs:

\begin{align*}
M_1 = \begin{pmatrix}
0 & 1 & 0 & 1\\
1 & 0 & 1 & 0\\
0 & 1 & 0 & 1\\
1 & 0 & 1 & 0\end{pmatrix}
\\
M_2 = \begin{pmatrix}
0 & 0 & 1 & 1\\
0 & 0 & 1 & 1\\
1 & 1 & 0 & 0\\
1 & 1 & 0 & 0\end{pmatrix}
\\
M_3 = \begin{pmatrix}
0 & 1 & 0 & 0\\
1 & 0 & 1 & 0\\
0 & 1 & 0 & 1\\
0 & 0 & 1 & 0\end{pmatrix}
\end{align*}

Graph represented by $M_1$ matrix has been evaluated by the quantum circuit on figure \ref{EF1}. Same way, figure \ref{EF2} shows the respective evaluations of circuits $M_2$ and $M_3$. All the values reached at measuring three bottom (control) qubits for $M_1$ and $M_2$ meet, so denoting they can be isomorphic graphs. On the contrary $M_3$ evaluation differs from any of the previous as the corresponding graph is isomorphic with none of them.

\[\begin{array}{ccc}
   \#(edges)  & \% Probability & \#(subgraphs) \\
    0 & 50.0 & 8 \\
    1 & 31.25 & 5 \\
    2 & 12.5 & 2 \\
    3 & 6.25 & 1 \\
\end{array}\]

\begin{algorithm}
    \caption{\label{alg1} Algorithm to codify the quantum circuit}
    Let $G=(V,E)$ be the graph, with $|V|=n$ and $|E| = m$
     
    $k = \lceil log_2 m \rceil$  
    
    Prepare the Oracle $M$ in the following way
    
    \For{every $(i,j) \in E, \, i < j$}{
        Add the gate Controlled $Z^{1/2^k}$ with control in qubit $i$ applied on qubit $j$
    }
    Prepare $n$ qubits initialized to $\dirac{+}$ for the Graph
    
    Prepare $k$ qubits initialized to $\dirac{+}$ for the Estimation
    
    \For{$i \in \{1,\ldots , k\}$}{ 
        Apply $2^{i-1}$ Controlled $M$ gates on all Graph qubits, with control in Estimation qubit $i$ 
    }
    Apply $QFT^\dagger$ on Estimation qubits
    
    Measure Estimation qubits
\end{algorithm}

This quantum oracle \ref{alg1} provides a powerful tool to be used onto graph isomorphism problem as identifying all the possible subgraphs of the one under consideration. This decomposition is indexed by the number of edges in each subgraph.

We have described the coding of a graph in a quantum circuit, and how to obtain its invariants by means of the table of subgraphs classified by probability, which offers us the number of edges. Now we must focus on the measure of it that will refer to the control qubits in the application of the phase estimation algorithm. First, we have that the observable to be measured does not depend on the order in which qubits are labelled, which is a necessary condition for an invariant. And second, in this case, we can use the tomography techniques described in \cite{Toth10} to obtain the measurement of the probabilities of the involved amplitudes.

\subsection{Correctness of the algorithm}

This section is devoted to proof that the method above described is correct.

Let us consider a non directed graph $G=(V, E)$, let us
assume that $V=\{0,\ldots, n-1\}$. For each edge $(i,j)\in E$ with $0\leq i<j<n$
let us consider the operator $A_{ij}^{\theta}$, with gates of precision $\theta$, defined as follows
\begin{displaymath}
  \Qcircuit @C=1em @R=.7em {
    \cdots & \cdots & \cdots\\
    \lstick{i} & \gate{P(\theta)} & \qw\\
    \cdots &  & \cdots\\
    \lstick{j} & \ctrl{-2} &\qw\\
    \cdots & \cdots & \cdots\\
  }
\end{displaymath}

This is formalised in the following definition

\begin{definition}
  Let $G=(V, E)$ with $V=\{0,\ldots, n-1\}$ be a non directed
  graph, $\theta \in \mathbb{R} \ s.t. \ 0 \leq \theta<2\pi$, and $(i,j)\in E$ with $0\leq i<j<n$. Let us consider the operator
\begin{displaymath}
\resizebox{\hsize}{!}{$
  A_{ij}^{\theta}=I^{\otimes i}\otimes \bigl(P(\theta)\otimes I^{\otimes j-i-1}\otimes
  \op{1}{1}+I^{\otimes j-i}\otimes \op{0}{0}\bigr)\otimes I^{\otimes n-j}
$}
\end{displaymath}
\end{definition}


Although we have not proved that the
operators $A_{ij}^{\theta}$ commute between them yet (this is consequence of
Lemma~\ref{lem:edge}), we have that the operator associated to the
graph appearing in Figure~\ref{F1} is given by the following
\begin{displaymath}
  U_{G}^{\theta}=\prod_{\substack{0\leq i<|V|,\\i<j<|V|,\\ (i,j)\in E}}A_{ij}^{\theta}
\end{displaymath}

Now let us consider a subgraph $G'=(V', E')$ of $G$. Let us consider
the vector
\begin{displaymath}
\resizebox{\hsize}{!}{$
  \Psi_{G'} = x_{0}\otimes \cdots x_{i}\otimes\cdots x_{j}\otimes\cdots
  x_{n-1}\quad\mathrm{where}\ x_{i}=
  \begin{cases}
    \ket{1} & \mathrm{if}\ i\in V'\\
    \ket{0} & \mathrm{if}\ i\not\in V'\\
  \end{cases}
$}
\end{displaymath}

\begin{lemma}\label{lem:edge}
  Let $G=(V,E)$ be a graph, $(i,j)\in E$ with $0\leq
  i<j<n$, \(0 < \theta < 2\pi\), and a subgraph $G'=(V', E')$. We have that 
  \begin{itemize}
  \item if $(i,j)\in E'$ then
    $A_{ij}^{\theta}\ket{\Psi_{G'}}=e^{i\theta}\ket{\Psi_{G'}}$
  \item if $(i,j)\not\in E'$ then $A_{ij}\ket{\Psi_{G'}}=\ket{\Psi_{G'}}$
  \end{itemize}
\end{lemma}

\begin{proof}
First, let us recall that \(P(\theta)\ket{0}=\ket{0}\) and
\(P(\theta)\ket{1}=e^{i\theta}\ket{1}\).

If we apply the operator
$A_{ij}^{\theta}$ to $\Psi_{G'}$ we obtain

\begin{multline*}
    A_{ij}^{\theta}\ket{\Psi_{G'}}=x_{0}\otimes \cdots x_{i-1}\otimes \\
    \Bigl(\bigl(P(\theta)x_{i}\otimes x_{i+1}\cdots \otimes
    x_{j-1}\otimes \op{1}{1}x_{j}\bigr) \\
    + \\
    \bigl(x_{i}\otimes x_{i+1}\cdots \otimes x_{j-1}\otimes
    \op{0}{0}x_{j}\bigr)\Bigr) \\
    \otimes x_{j+1}\otimes\cdots x_{n-1}
\end{multline*}

Let us consider the following cases:
\begin{itemize}
\item $j\not\in V'$, then \(x_{j}=\ket{0}\). In this case
  \(\op{1}{1}x_{j}=\op{1}{1}\ket{0}=0\)
  and \(\op{0}{0}x_{j}=\op{0}{0}\ket{0}=\ket{0}\).
  Therefore
  \begin{gather*}
    \Bigl(\bigl(\overbrace{P(\theta)x_{i}\otimes x_{i+1}\cdots\otimes
    x_{j-1}\otimes \overbrace{\op{1}{1}x_{j}}^{=0}}^{=0}\bigr) \\
    + \\
    \bigl(x_{i}\otimes x_{i+1}\bigl(\otimes x_{j-1}\otimes
    \overbrace{\op{0}{0}x_{j}}^{=x_{j}}\bigr)\Bigr) \\
    = \\
    x_{i}\otimes x_{i+1}\cdots\otimes x_{j-1}\otimes x_{j}
  \end{gather*}
\item \(i\not\in V'\) and \(j\in V'\). In this case we have
  $x_{i}=\ket{0}$, $x_{i}=\ket{1}$,
  \(P(\theta)x_{i}=P(\theta)\ket{0}=\ket{0}=x_{i}\),
  \(\op{1}{1}x_{j}=\op{1}{1}\ket{1}=\ket{1}=x_{j}\),
  and \(\op{0}{0}x_{j}=\op{0}{0}\ket{1}=0\). Therefore,
  \begin{gather*}
      \Bigl(\bigl(\overbrace{P(\theta)x_{i}}^{=x_{i}}\otimes x_{i+1}\cdots\otimes
      x_{j-1}\otimes \overbrace{\op{1}{1}x_{j}}^{=x_{j}}\bigr) \\
      + \\
      \overbrace{\bigl(x_{i}\otimes x_{i+1}\bigl(\otimes x_{j-1}\otimes
      \overbrace{\op{0}{0}x_{j}}^{=0}\bigr)}^{=0}\Bigr) \\
      = \\
      x_{i}\otimes x_{i+1}\cdots\otimes x_{j-1}\otimes x_{j}
  \end{gather*}
\item \((i,j)\in E'\). This case is similar to the previous one. But
  $x_{i}=\ket{1}$ and \(P(\theta)\ket{1}=e^{i\theta}\ket{1}\). Therefore,
  \begin{gather*}
      \Bigl(\bigl(\overbrace{P(\theta)x_{i}}^{=e^{i\theta}x_{i}}\otimes x_{i+1}\cdots\otimes
      x_{j-1}\otimes \overbrace{\op{1}{1}x_{j}}^{=x_{j}}\bigr) \\
      + \\
      \overbrace{\bigl(x_{i}\otimes x_{i+1}\bigl(\otimes x_{j-1}\otimes
      \overbrace{\op{0}{0}x_{j}}^{=0}\bigr)}^{=0}\Bigr) \\
      = \\
      e^{i\theta}x_{i}\otimes x_{i+1}\cdots\otimes x_{j-1}\otimes x_{j}
  \end{gather*}
  Since \(e^{i\theta}\) is a scalar,  we obtain $A_{ij}^{\theta}\ket{\Psi_{G'}}=e^{i\theta}\ket{\Psi_{G'}}$
\end{itemize}
\end{proof}

A clear consequence of the previous Proposition is that the operators $A_{ij}^{\theta}$ commute,
so the following definition is correct,
\begin{definition}
  Let $G=(V, E)$ be a non directed
  graph and $0<\theta<2\pi$. We define
  \begin{displaymath}
    U_{G}^{\theta}=\prod_{\substack{0\leq i<j<|V| \\ (i,j)\in E}}A_{ij}^{\theta}
  \end{displaymath}
\end{definition}

\begin{proposition}\label{prop:subgraph}
Let us consider $G=(V,E)$ a graph, and $(i,j)\in E$ with $0\leq
i<j<n$, $1<\theta<2\pi$, and a subgraph $G'=(V', E')$. Then
\begin{equation}\label{eq:graph}
  U_{G}^{\theta}\ket{\Psi_{G'}}=e^{i\theta|E'|}\ket{\Psi_{G'}}
\end{equation}
\end{proposition}
\begin{proof}
 This is a direct consequence of the previous Lemma~\ref{lem:edge}.
\end{proof}

In order to obtain $|E'|$ we can apply the Quantum Phase Estimation (QPE in short)
circuit. But first, we have to rewrite Equation~\ref{eq:graph}. We have not given a proper value of \(\theta\) yet but applying
the QPE circuit is required, therefore, the value of \(\theta\) is restricted. First, we
are going to apply the QPE circuit to estimate the number of
subgraphs that a given graph \(G=(V,E)\) has.
Let us consider $e=\lceil \log_{2}{|E|}\rceil$ and
\begin{displaymath}
  e' =
  \begin{cases}
    e & \ \text{if} \ |E| \ \text{is not a power of 2}\\
    e+1 & \ \text{if} \ |E| \ \text{is a power of 2}\\
  \end{cases}
\end{displaymath}
Therefore for \(\theta=\frac{2\pi}{2^{e'}}\), we consider the QPE circuit for $U_{G}^{\theta}$ and $e$
qubits for estimating the number of edges of a subgraph $G'\subseteq G$.
Let us start on Equation~\eqref{eq:graph}.
\begin{multline}
  U_{G}^{\theta}\ket{\Psi_{G'}}=e^{i\theta|E'|}\ket{\Psi_{G'}} \\
  = e^{2\pi i\frac{\theta|E'|}{2\pi}}\ket{\Psi_{G'}}=e^{2\pi{}i\frac{|E'|}{2^{e'}}}\ket{\Psi_{G'}}
\end{multline}

Now we can use the QPE estimation circuit to estimate $\frac{|E'|}{2^{e'}}$.

Let us consider the Quantum Phase Estimation \(QPE_{G}\) operator built from
the operator $U_{G}^{\theta}$. As we have seen, all the elements of the canonical base
correspond to a subgraph and vice-versa. If we apply the operator
\(QPE_{G}\) to $\Psi_{G'}$ we obtain:

\begin{equation}
  QPE_{G}\ket{\Psi_{G'}}=\ket{\chi_{\Psi_{G'}}\otimes \Psi_{G'}}
\end{equation}

Where $\chi_{\Psi_{G'}}$ is the binary representation of
\(\frac{|E'|}{2^{e'}}\). Since \(\frac{|E'|}{2^{e'}}<1\) and \(|E'|=2^{e'}\frac{|E'|}{2^{e'}}\).

Besides, $\chi_{\Psi_{G'}}$ is also the binary representation of \(|E'|\) as a natural number.

Our algorithm can be summarised as
\begin{displaymath}
    \Qcircuit @C=1em @R=1.2em {
    \lstick{\ket{0}} & {/}\qw &\qw                &\qw& \multigate{1}{QPE_{G}}        &\qw & \qw  \\
    \lstick{\ket{0}} & {/}\qw & \gate{H^{\otimes n}}&\qw& \ghost{QPE_{G}} &\qw & \meter \\
  }
\end{displaymath}

Mathematically it is \(QPE_{G}(I^{e}\oplus H^{\otimes |V|})\), where $e$
is the number of bits of the estimation (as indicated before).

\begin{theorem}
  Let $G=(V,E)$ be a non directed graph.
  Then
  \begin{multline*}
    QPE_{G}(I^{e}\oplus H^{\otimes |V|})\ket{0^{e+|V|}} \\
    = \sum_{
        \substack{
        x\in \{0,1\}^{e},\\ G'=(V',E'),\\ x=|E'|
      }
    } 2^{-\frac{n}{2}}\ket{\tilde{x}\otimes\Psi_{G'}}
  \end{multline*}
\end{theorem}

\begin{proof}
  \begin{equation}
    \begin{split}
      QPE_{G}\bigl(I^{e}\otimes H^{\otimes n}\bigr)\ket{0^{\otimes e}}\otimes\ket{0^{\otimes n}} = \\
      QPE_{G}\ket{0^{\otimes e}}\otimes\ket{\Bigl(2^{-\frac{n}{2}}\sum_{G'\subseteq G}\Psi_{G'} \Bigr)} =\\
      2^{-\frac{n}{2}}\sum_{G'\subseteq G}QPE_{G}\ket{0^{\otimes e}}\otimes\ket{\Psi_{G'}}=\\
      2^{-\frac{n}{2}}\sum_{G'\subseteq G}\ket{\chi_{\Psi_{G'}}\otimes\Psi_{G'}}=\\
      =\sum_{x\in
        {0,1}^{e}}\sum_{x=\chi_{\Psi_{G'}}}2^{-\frac{n}{2}}\ket{\tilde{x}\otimes\Psi_{G'}}
    \end{split}
  \end{equation}

  First of all, if $x=\chi_{\Psi_{G'}}$ is the $l-bit$ binary
  representation of a number between 0 and 1. So $2^{l}x$ is an
  integer whose representation in base $2$ is also $x$, but now
  interpreted as an integer. Let us recall that
  $\chi_{\Psi_{G'}}=\frac{\theta|E'|}{2\pi}$ and
  $\theta=\frac{2\pi}{2^{l}}$. So
  $\chi_{\Psi_{G'}}=\frac{|E'|}{2^{l}}$, so $2^{l}x=|E'|$. If we
  interpret $x$ as an integer, $x=|E'|$.  Therefore, the last line of
  the previous equation can be rewritten as
  \begin{displaymath}
    \begin{split}
      \sum_{x\in \{0,1\}^{e}} \sum_{\substack{G'=(V',E'),\\ x=|E'|}}
      2^{-\frac{n}{2}}\ket{\tilde{x}\otimes\Psi_{G'}}=\\
      \sum_{ \substack{ x\in \{0,1\}^{e},\\ G'=(V',E'),\\ x=|E'| } }
      2^{-\frac{n}{2}}\ket{\tilde{x}\otimes\Psi_{G'}}
    \end{split}
  \end{displaymath}
\end{proof}

Finally we can measure the first e qubits of the final state
\begin{displaymath}
      \sum_{ \substack{ x\in \{0,1\}^{e},\\ G'=(V',E'),\\ x=|E'| } }
      2^{-\frac{n}{2}}\ket{\tilde{x}\otimes\Psi_{G'}}
\end{displaymath}
All elements in the last term share the coefficient \(2^{-\frac{n}{2}}\), thus the probability of computing $x$ is
  \begin{displaymath}
    \begin{split}
      p(x) = (2^{-\frac{n}{2}})^{2} \Bigl|\bigl\{\Psi_{G'}\ |\ G'=(V',E'), x=|E'| \bigr\}\Bigr|=\\
      2^{-n}\Bigl|\bigl\{\Psi_{G'}\ |\ G'=(V',E'), x=|E'| \bigr\}\Bigr|
  \end{split}
\end{displaymath}

\subsection{Discussion}

Although there is no exhaustivity on the presented codification, this is a very strong and useful invariant as the number of subgraphs in a $n$-nodes graph is $2^n$ and we provide a way to classify all of them in polynomial time over the number of edges of that graph.

From the computational complexity point of view we wonder whether it can perform within polynomial time order. We start focusing on the amount of control qubits required as well as the number of times we need invoking the oracle which models the graph under investigation. Both issues lay on the accuracy of the quantum gate $Z^k$ appropriate for the case, where $k=1/2^{m} \ , m \in\mathbb N$ which implies the need of $m-1$ control qubits besides a number of oracle calls performing of $2^{m-1}-1$ times.

For the particular case of the $C_4$ graph we are using as example all along the present piece of research, the corresponding quantum gate is $T=Z^{1/4}$, thus $3$ control qubits are used, and the oracle is being called $2^3 -1 = 7$ times.

It is immediate that the value for $k$, following the pattern $1/2^{m}$, fully determines the previous two key elements in complexity estimation. Previously, we have elaborated how the number of edges in the graph imposes a lower bound for $k$ so that the angle $\pi/k$ multiplied by the number of edges must be smaller than $2 \pi$ 

It is straightforward that the amount of control qubits belongs to the order of the logarithm (base 2) of the total number of edges in the graph, besides, the number of oracle requests meets the order of the number of edges. As in the worst case the number of edges is $n\cdot(n-1)$, where $n$ is the number of nodes of the graph, we can estate that this algorithm takes polynomial time on the number of nodes of the graph.

Of course, this cannot be interpreted as full complexity of algorithm was polynomial, because computing the computational cost of measuring is still pending.

We want to know the probabilities of each control qubit. For this sake, as this operation is just performed once over a register set to $\ket{H^{\otimes n} \otimes H^{\otimes k}}$  ($n + k$ qubits) thus the algorithm will output  

\[
  Alg(\ket{H^{\otimes n} \otimes H^{\otimes k}}) = QPE_{G}\ket{\Psi_{G'}}=\ket{\chi_{\Psi_{G'}}\otimes \Psi_{G'}}
\]

Then, we need the {\em partial trace} on the second component from the register so obtained.

\[ {\rm tr_{\chi_{\Psi_{G'}}}} \ket{\chi_{\Psi_{G'}}\otimes \Psi_{G'}}
\bra{\chi_{\Psi_{G'}}\otimes \Psi_{G'}}
\]

But the calculation of the partial trace from the state of the system requires the evaluation of the amplitudes of it that, according with Postulate 3 of Quantum Mechanics \cite{nielsen00}, cannot be performed. Therefore, on practice, estimating these values necessarily involves either repeating executions and measurements several times (Montecarlo fashioned) or, some other tomography techniques. Both options carry a need of exponential time when looking for $m$ precision digits for the correct value. 

\section{Empirical accuracy and limits}

For the sake of briefly, and roughly, illustrating the accuracy of our invariant proposal the table \ref{tab:invariant-comparison} summarises the figures of accuracy for all graphs up to seven nodes.

\begin{table*}[htbp]
\begin{tabular}{lccccccc}
\hline \hline
Number of nodes                              & 1 & 2 & 3 & 4  & 5     & 6      & 7         \\ \hline
Number of graphs                             & 1 & 2 & 8 & 64 & 1,024 & 32,768 & 2,097,152 \\
Number of non isomorphic graphs              & 1 & 2 & 4 & 11 & 34    & 156    & 1,044     \\
Number of quantum invariant different graphs & 1 & 2 & 4 & 11 & 34    & 156    & 1,021     \\
Number of different eigenspectra             & 1 & 2 & 4 & 11 & 33    & 151    & 988       \\ \hline \hline
\end{tabular}
\caption{Invariant accuracy up to 7 nodes}
\label{tab:invariant-comparison}
\end{table*}

\subsection{Petersen and Pentagonal Prism graphs}

\begin{figure}[htbp]
    \centering
    \includegraphics[width=0.7\hsize]{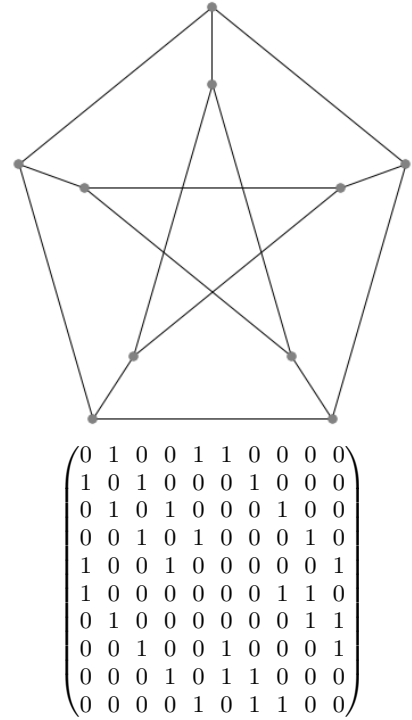}
    $
    \begin{pmatrix}
        0 & 1 & 0 & 0 & 1 & 1 & 0 & 0 & 0 & 0\\
        1 & 0 & 1 & 0 & 0 & 0 & 1 & 0 & 0 & 0\\
        0 & 1 & 0 & 1 & 0 & 0 & 0 & 1 & 0 & 0\\
        0 & 0 & 1 & 0 & 1 & 0 & 0 & 0 & 1 & 0\\
        1 & 0 & 0 & 1 & 0 & 0 & 0 & 0 & 0 & 1\\
        1 & 0 & 0 & 0 & 0 & 0 & 0 & 1 & 1 & 0\\
        0 & 1 & 0 & 0 & 0 & 0 & 0 & 0 & 1 & 1\\
        0 & 0 & 1 & 0 & 0 & 1 & 0 & 0 & 0 & 1\\
        0 & 0 & 0 & 1 & 0 & 1 & 1 & 0 & 0 & 0\\
        0 & 0 & 0 & 0 & 1 & 0 & 1 & 1 & 0 & 0
    \end{pmatrix}
    $
    \caption{Petersen Prism graph}
    \label{fig:petersen}
\end{figure}

\begin{figure}[htbp]
    \centering
    \includegraphics[width=0.7\hsize]{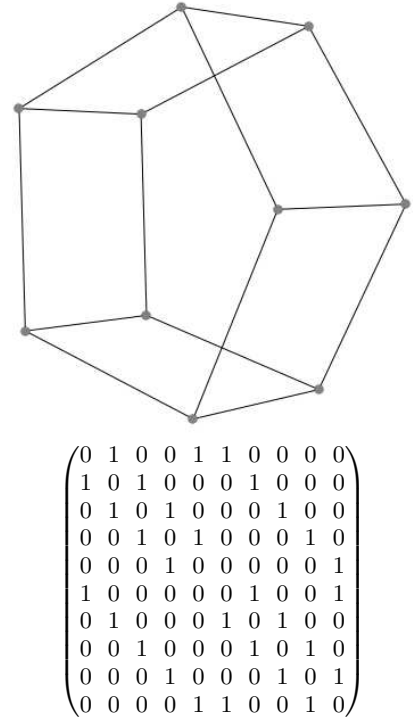}
    $
    \begin{pmatrix}
        0 & 1 & 0 & 0 & 1 & 1 & 0 & 0 & 0 & 0\\
        1 & 0 & 1 & 0 & 0 & 0 & 1 & 0 & 0 & 0\\
        0 & 1 & 0 & 1 & 0 & 0 & 0 & 1 & 0 & 0\\
        0 & 0 & 1 & 0 & 1 & 0 & 0 & 0 & 1 & 0\\
        0 & 0 & 0 & 1 & 0 & 0 & 0 & 0 & 0 & 1\\
        1 & 0 & 0 & 0 & 0 & 0 & 1 & 0 & 0 & 1\\
        0 & 1 & 0 & 0 & 0 & 1 & 0 & 1 & 0 & 0\\
        0 & 0 & 1 & 0 & 0 & 0 & 1 & 0 & 1 & 0\\
        0 & 0 & 0 & 1 & 0 & 0 & 0 & 1 & 0 & 1\\
        0 & 0 & 0 & 0 & 1 & 1 & 0 & 0 & 1 & 0
    \end{pmatrix}
    $
    \caption{Pentagonal Prism graph}
    \label{fig:pentagonal}
\end{figure}

We will work over a couple of well known 10 nodes, 3 edges per node, regular graphs as Petersen and Pentagonal Prism graph. Their adjacency matrices are placed over their respective graphs as shown in figures \ref{fig:petersen} and \ref{fig:pentagonal}. We number the nodes in Petersen graph by inner pentagon with 1 on the top and clockwise, then 6 to the one on top of the outer pentagon and clockwise too. Same way of numbering for Pentagonal Prism one, no matter which is consider as inner/outer.

Let us codify each of the graphs following the way previously described for being taken as oracle for further analysis. The quantum gate required for this is $Z^{1/8}$ which holds the required accuracy as far as rotations amplitudes of $\pi/8$ allows up to 15 times before reaching $2\pi$ so corresponding to the base vector $\dirac{1111111111}$ which represents the whole graph. Its amplitude equals $1/32\cdot e ^{i \, 15/8 \pi }$.

We require 4 ($=Roof(\log_2(15))$) control qubits and 15 times oracle calls in order to phase estimating. Once all this stuff has been charged on Quirk quantum simulator the outcome is summarized in the table \ref{tab:pet-results} so classifying all their subgraphs.

\begin{table*}[htbp]
\begin{tabular}{ccccc}
          & \multicolumn{2}{c}{Petersen}   & \multicolumn{2}{c}{Pentagonal Prism} \\
\#(edges) & \% Probability & \#(subgraphs) & \% Probability    & \#(subgraphs)    \\
    0 & 7.42 & 76   & 7.91 & 81\\
    1 & 13.18 & 135 & 12.21 & 125\\
    2 & 16.13 & 165 & 15.14 & 155\\
    3 & 13.18 & 135 & 17.58 & 180\\
    4 & 17.58 & 180 & 12.21 & 125\\
    5 & 8.50 & 87  & 12.40 & 127\\
    6 & 9.77 & 100 & 7.81 & 80\\
    7 & 5.86 & 60  & 6.35 & 65\\
    8 & 2.93 & 30  & 2.93 & 30\\
    9 & 2.93 & 30  & 2.93 & 30\\
    10 & 1.46 & 15 & 1.46 & 15\\
    11 & 0.00 & 0  & 0.00 & 0 \\
    12 & 0.97 & 10 & 0.97 & 10\\
    13 & 0.00 & 0  & 0.00 & 0 \\
    14 & 0.00 & 0  & 0.00 & 0 \\
    15 & 0.10 & 1  & 0.10 & 1\\
\end{tabular}
\caption{Petersen and Pentagonal Prism results}
\label{tab:pet-results}
\end{table*}

This table shows that both graphs are not isomorphic. If we only took under consideration those subgraphs having 8 or more edges they would seem isomorphic graphs, but taking a deep enough sight of the table it is easy to identify their differences for subgraphs of smaller number of edges. In fact, there are 76 independent subgraphs (with no edges) for Petersen graph meanwhile for Pentagonal Prism one there are 81 of these sort of subgraphs (empty and 10 single node graphs included in both cases).

Even more, if we will be able to measure the amplitude of $\dirac{00001010001010}$ in the quantum circuit for Pentagonal Prism graph we will get $0$ even when it is one of those included within the 81 previously mentioned (those without any edge), and, as it is the largest number of nodes for this case, we have that subgraph 
$\{2, 4, 8, 10\}$ corresponds to the maximal independent set.

\subsection{Counterexample}

This proposal is just an invariant, that does not characterise the isomorphism over graphs.

\[\begin{pmatrix}0 & 1 & 0 & 0 & 0 & 1 & 1\\
1 & 0 & 1 & 0 & 0 & 0 & 1\\
0 & 1 & 0 & 1 & 0 & 0 & 0\\
0 & 0 & 1 & 0 & 1 & 0 & 0\\
0 & 0 & 0 & 1 & 0 & 1 & 0\\
1 & 0 & 0 & 0 & 1 & 0 & 0\\
1 & 1 & 0 & 0 & 0 & 0 & 0\end{pmatrix}
\hspace{2em}
\begin{pmatrix}0 & 1 & 0 & 1 & 1 & 0 & 0\\
1 & 0 & 1 & 0 & 0 & 0 & 0\\
0 & 1 & 0 & 1 & 0 & 0 & 0\\
1 & 0 & 1 & 0 & 0 & 0 & 0\\
1 & 0 & 0 & 0 & 0 & 1 & 1\\
0 & 0 & 0 & 0 & 1 & 0 & 1\\
0 & 0 & 0 & 0 & 1 & 1 & 0\end{pmatrix}\]

\begin{figure}[htbp]
    \centering
    \includegraphics[width=5cm]{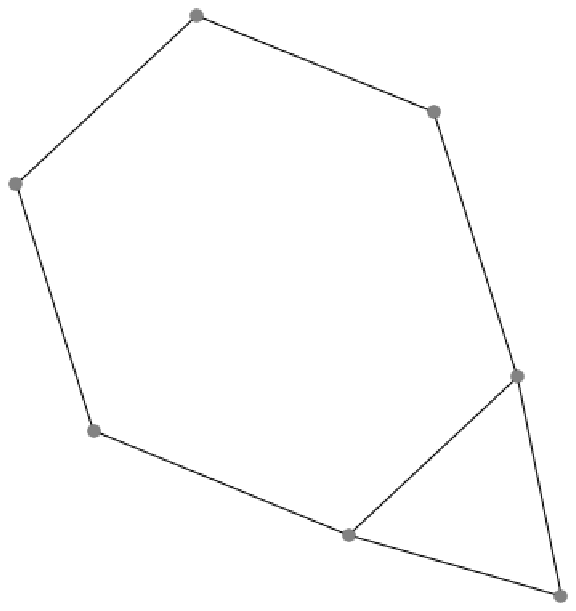}
    \includegraphics[width=5cm]{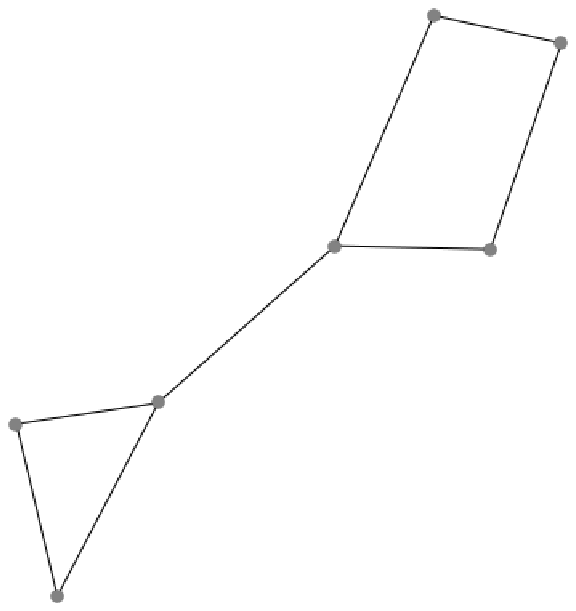}
  \caption{Two 7 nodes non isomorphic graphs with same invariant}
  \label{Counter}
\end{figure}

\begin{table*}[htbp]
\begin{tabular}{ccccc}
          & \multicolumn{2}{c}{G1}         & \multicolumn{2}{c}{G2}         \\
\#(edges) & \% Probability & \#(subgraphs) & \% Probability & \#(subgraphs) \\
    0 & 20.31 & 26   & 20.31 & 26\\
    1 & 25.78 & 33 & 25.78 & 33\\
    2 & 21.09 & 27 & 21.09 & 27\\
    3 & 14.06 & 18 & 14.06 & 18\\
    4 & 10.16 & 13 & 10.16 & 13\\
    5 & 3.91 & 5  & 3.91 & 5\\
    6 & 3.91 & 5  & 3.91 & 5\\
    7 & 0.00 & 0  & 0.00 & 0\\
    8 & 2.93 & 1  & 2.93 & 1\\
\end{tabular}
\caption{G1 and G2 graph results}
\label{tab:g12-results}
\end{table*}

\section{Conclusions and future work}
In this work we have studied some interesting questions related to invariants on the graph isomorphism problem from the quantum information perspective.

To begin with, we have presented a new way for encoding non directed graphs of $n$ nodes, by means of quantum gates on $n$ qubits, in such a way that we can establish a relationship between the adjacency characteristics of the nodes of the graph, and the angles of each amplitude of the quantum state of the system. The latter corresponds to the subset of nodes of the graph identified by 1 in the sequence of bits that defines each amplitude.

Once a graph has been encoded in the described fashion, applying the phase estimation algorithm outputs valuable information about all subgraphs of the original one, classified by the number of edges they contain. We can note that there is an exponential number of subgraphs, so we have provided an algorithm that can sort out that exponential number of graphs within polynomial time  (apart from the task of measuring the system).

In fact, such piece of information provides us with a powerful invariant on graph isomorphism. Some examples are provided with the aim of illustrating its distinguishing power. However, it does not discriminate all graphs as in counterexample section is shown.

We still know neither under what circumstances non-isomorphic graphs can share the same classification table of subgraphs by number of edges, nor, whether graphs considered hard to be classified by classical algorithms could be properly discriminated (or not) by our algorithm, e.g., strong regular graphs.

\section*{Acknowledgements}
This work has been supported by the Spanish MINECO/FEDER project AwESOMe (PID2021-122215NB-C31) and the Region of Madrid project FORTE-CM (S2018/TCS-4314) co-funded by EIE Funds of the European Union and the Qsimov Quantum Computing project `Elaboracion de una plataforma de computacion cuantica, y actividades formativas' (220426UCTR) 

\nocite{*}

\bibliography{apssamp}

\end{document}